\documentclass[twocolumn,showpacs,aps,superscriptaddress]{revtex4-2}
\usepackage{color}
\usepackage{soul}
\usepackage{lipsum}  
\usepackage{graphicx}% Include figure files
\usepackage{dcolumn}% Align table columns on decimal point
\usepackage{bm}% bold math
\usepackage{hyperref}
\usepackage{amsmath}
\usepackage{mathtools}
\usepackage[export]{adjustbox}
\begin{document}

\preprint{APS/123-QED}

\title{Rank-dependent optimal resetting in multiparticle search}
%\thanks{A footnote to the article title}%

\author{Ron Vatash}
\affiliation{The Raymond and Beverley School of Chemistry, Tel Aviv University, Tel Aviv 6997801, Israel.}
\author{Eden Goldfarb}
\affiliation{The Raymond and Beverley School of Physics \& Astronomy, Tel Aviv University, Tel Aviv 6997801, Israel.}
\author{Vladimir Yu. Rudyak}
\affiliation{The Raymond and Beverley School of Physics \& Astronomy, Tel Aviv University, Tel Aviv 6997801, Israel.}
\author{Yael Roichman}
\affiliation{The Raymond and Beverley School of Chemistry, Tel Aviv University, Tel Aviv 6997801, Israel.}
\affiliation{The Raymond and Beverley School of Physics \& Astronomy, Tel Aviv University, Tel Aviv 6997801, Israel.}

\date{\today}

\begin{abstract}
In many soft-matter and biological systems, task completion relies on the
cumulative arrival of multiple searchers rather than the speed of a single pioneer. The completion kinetics are therefore set not only by the first arrival, but by the full ordered sequence of first-passage times. Here, we determine how stochastic resetting optimizes these ordered arrivals for all arrival ranks. We construct an exact finite-$N$ reference for non-interacting Brownian searchers and obtain the mean ordered first-passage time $\langle T_{(k)} \rangle$ and its optimal resetting rate $r_k^*$. For searchers with identical initial conditions, $r_k^*$ increases monotonically with arrival rank and, with increasing population size, approaches the known large-$N$ quantile limit where a finite optimum appears only above a critical rank fraction $\phi_c \simeq 0.412$. Spatial heterogeneity qualitatively reorganizes this sequence, shifting its maximum from late to early ranks even without particle interactions. We then compare this baseline with Brownian colloid experiments, interacting active Brownian particles, and a collective autochemotactic search with persistent environmental memory. Across these systems, sensitivity to resetting increases strongly with arrival rank, while deviations from appropriate non-interacting references reveal the influence of direct interactions, finite return overhead, and environmental memory. Our results show that optimal resetting in multiparticle search is governed by the required completion rank and must be evaluated relative to protocol- and geometry-matched baselines.
\end{abstract}

\maketitle

%****************
%intro
%***************
\section{Introduction}
% --- INTRODUCTION (First Paragraph) ---
While standard first-passage theory evaluates search efficiency by the fastest individual,
many physical and biochemical processes are triggered only when a threshold number
of constituents reaches or binds a target. For example, diffusion-controlled
reactions may require a prescribed number $k$ out of $N$ particles to bind before a
downstream event initiates~\cite{GrebenkovKumar2022}. More generally, collective
capture, recruitment, and assembly processes depend on the accumulation of multiple
successful searchers rather than on the fastest one alone. In such cases, a single
first-passage time fails to capture the underlying completion kinetics.

This distinction is naturally expressed through the order statistics of
first-passage times. If $T_i$ denotes the first time at which searcher $i$
reaches a target, we order the individual passage times as
\begin{equation}
	T_{(1)} \leq T_{(2)} \leq \cdots \leq T_{(N)},
\end{equation}
where $T_{(k)}$ is the time at which the $k$th distinct member of the
population reaches the target. The relevant completion criterion can therefore
range from the fastest arrival, $T_{(1)}$, through an intermediate threshold
$T_{(k)}$, to the arrival of the complete population, $T_{(N)}$. Ordered
first-passage times of independent diffusing particles have a long history in
statistical physics~\cite{WeissShulerLindenberg1983,YusteLindenberg1996}, and
have recently been emphasized in multiparticle reaction and binding
kinetics~\cite{GrebenkovKumar2022}.

Many physical search and assembly processes also contain mechanisms that
terminate unsuccessful excursions and initiate new attempts. While distinct
from literal ordered first passage, these dynamics motivate the renewal
mechanism studied here. Incomplete molecular assemblies disassemble and
reform, as observed in pore and viral-capsid assembly~\cite{Thompson2011,Zlotnick2007},
while microtubule search and capture relies on stochastic switching between
growth and rapid shrinkage~\cite{MitchisonKirschner1984}. In the latter case,
tuning the growth and catastrophe rates optimizes the search time~\cite{HolyLeibler1994}.
The common element is the abandonment of unsuccessful trajectories or
configurations followed by a fresh attempt.

Stochastic resetting provides a minimal framework for isolating this
mechanism. In its simplest form, a search trajectory returns to a
prescribed state at Poisson-distributed times. For a diffusive searcher,
resetting truncates long unsuccessful excursions and produces a finite
resetting rate that minimizes the mean first-passage time~\cite{EvansMajumdar2011}.
This result has motivated a broad theory of restarted first passage~\cite{Reuveni2016,PalReuveni2017,EvansMajumdarSchehr2020}
and extensions to persistent and active dynamics~\cite{EvansMajumdar2018,KumarSadekarBasu2020,BaoucheKurzthaler2025}.

For a given search criterion, we define the optimal resetting rate \(r^*\), as the rate that minimizes the corresponding mean passage time. 
For searcher populations, most optimization studies have focused on the
first arrival. Biroli, Majumdar, and Schehr studied $T_{(1)}$ for Brownian
searchers under independent and common resetting protocols and identified a
protocol-dependent critical population size above which resetting ceases to
benefit the fastest arrival (\(r^*\to0\)) ~\cite{BiroliMajumdarSchehr2023}. 
In the opposite limit, $N \to \infty$ at fixed rank fraction $\phi = k/N$, the $k$th
ordered arrival of independent identically distributed searchers converges to
the $\phi$-quantile of the single-particle first-passage distribution. For
one-dimensional Brownian first passage, Belan showed that resetting lowers
this quantile only above $\phi_c \simeq 0.4123$~\cite{Belan2020}. Recent work
has also begun to compare alternative collective completion criteria,
including first and median group-hitting times under state-dependent group
resetting~\cite{LeeYangLizana2026}.

However, a finite-$N$ description spanning the complete sequence
of arrival ranks remains lacking, particularly regarding how the corresponding
optimal resetting rates are altered when the assumptions of identical
independent searchers are relaxed. At finite $N$, arrival ranks remain discrete.
Different reset distances make searchers non-identically distributed, a shared
reset clock induces statistical correlations even without physical interactions,
and direct interactions, active persistence, finite return protocols, or
environmental memory can further couple search dynamics. Separating these
effects therefore requires both a tractable finite-$N$ reference and
system-specific non-interacting controls.

Here, we first construct an exact finite-$N$ reference for non-interacting
Brownian searchers under independent resetting. By combining the restarted
single-particle survival probability with exact order statistics, we obtain
$\langle T_{(k)}\rangle$ and its optimal rate $r_k^*$ for arbitrary rank and
arbitrary initial distance configurations. For searchers with identical
starting positions, $r_k^*$ increases monotonically with rank and approaches
the known large-$N$ quantile limit~\cite{Belan2020}. Spatial heterogeneity
alone, however, qualitatively reorganizes this sequence, shifting the maximum
of $r_k^*$ toward earlier ranks even without particle interactions. We further
show that changing from independent to simultaneous global resetting modifies
the optimal-rate sequence even for otherwise non-interacting searchers.

We then examine three physical search systems: Brownian colloids under global
resetting~\cite{VatashRoichman2025}, interacting active Brownian particles
under local resetting, and a collective autochemotactic search in which
previous trajectories are stored in a persistent chemical
field~\cite{Rudyak2025,RudyakRoichman2026}. For each physical system, we construct a corresponding non-interacting
reference that retains the same geometry, resetting rules, and underlying
single-particle motion, while removing the collective coupling mechanisms.
Comparing each system with its corresponding reference allows the effects of
these additional physical couplings to be distinguished from the substantial
rank dependence already generated by ordered statistics, geometry, and
resetting protocol.

% ---------------------------------------------------------------------------
% RESULTS
% ---------------------------------------------------------------------------

\section{Results}

\subsection{Ordered arrivals under stochastic resetting}

\begin{figure}[h]
	\centering
	\includegraphics[width=0.9\linewidth]{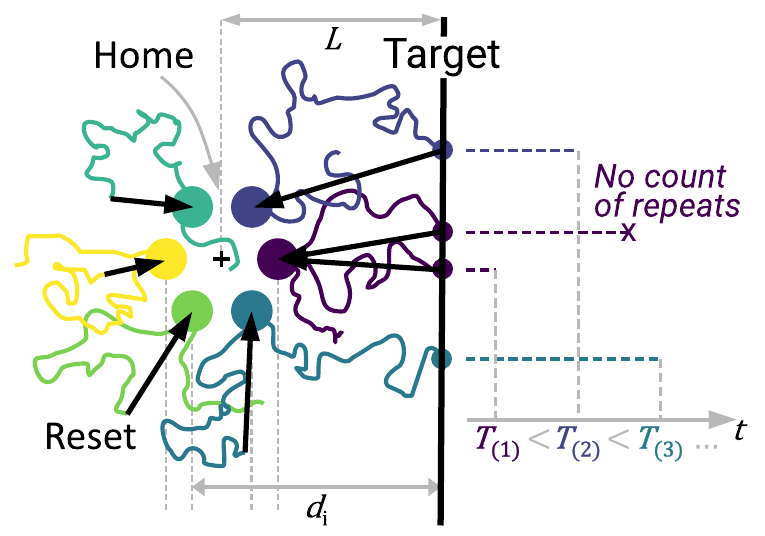}
	\caption{\textbf{Ordered first passage under stochastic resetting.}
		$N$ searchers begin from prescribed initial positions and search for a
		common target. During the search, stochastic resetting returns a particle
		to its reset position. The first-passage times $T_i$ of the distinct
		searchers are ordered as $T_{(1)}\leq\cdots\leq T_{(N)}$.
		The resetting rate minimizing the corresponding mean ordered first-passage
		time $\langle T_{(k)}\rangle$ is denoted $r_k^*$.}
	\label{fig:scheme}
\end{figure}

We first construct the simplest non-interacting reference for this work.
As illustrated schematically in Fig.~\ref{fig:scheme}, we consider $N$
Brownian searchers with diffusion coefficient $D$. The target is located at
distance $L$ from the center of the initial configuration, while particle $i$
starts at a distance $d_i$ from the target. The set of initial target
distances is therefore $\{d_i\}_{i=1}^{N}$. Each particle returns to its
initial position via an independent Poisson resetting process at rate $r$.
Because the searchers neither interact nor share a reset clock, their
trajectories and first-passage times remain statistically independent.

For a searcher initially at distance $d_i$, let $Q_r(t|d_i)$ denote the
survival probability under stochastic resetting. Its Laplace transform obeys the
standard renewal relation~\cite{EvansMajumdar2011},
\begin{equation}
	\widetilde Q_r(s|d_i)
	=
	\frac{\widetilde Q_0(s+r|d_i)}
	{1-r\widetilde Q_0(s+r|d_i)},
	\label{eq:renewal}
\end{equation}
where, for one-dimensional diffusion toward an absorbing boundary,
\begin{equation}
	\widetilde Q_0(s|d_i)
	=
	\frac{1-\exp[-d_i\sqrt{s/D}]}{s}.
	\label{eq:laplace}
\end{equation}
The corresponding restarted first-passage cumulative distribution function is
$F_i(t) = 1 - Q_r(t|d_i)$.

For independent but non-identically distributed searchers, the condition
$T_{(k)} > t$ requires that fewer than $k$ particles reach the target by time $t$.
Thus,
\begin{equation}
	\Pr[T_{(k)}>t]
	=
	\sum_{m=0}^{k-1}
	\sum_{|S|=m}
	\prod_{i\in S}F_i(t)
	\prod_{j\notin S}\left[1-F_j(t)\right],
	\label{eq:probTk}
\end{equation}
and
\begin{equation}
	\left\langle T_{(k)}\right\rangle
	=
	\int_{0}^{\infty}
	\Pr[T_{(k)}>t]\,\mathrm{d}t .
	\label{eq:meanTk}
\end{equation}
Equations~\eqref{eq:probTk}--\eqref{eq:meanTk} yield an exact finite-$N$
formulation for arbitrary initial distance configurations. This framework allows the effects of arrival rank and spatial geometry to be examined before introducing physical interactions or reset-induced correlations.

When initial positions are identical ($d_i = d$), the first-passage times become
independent and identically distributed (i.i.d.), reducing Eq.~\eqref{eq:probTk}
to classical binomial order statistics.

\subsection{Rank and reset geometry determine optimal resetting}

We begin with $N$ identical Brownian searchers starting at distance $d$ from
the target and resetting independently to their initial positions. In this
homogeneous baseline, the dynamics depend on the resetting rate via the
dimensionless combination $rd^2/D$, establishing a clean setting to isolate
arrival rank from spatial asymmetry, correlations, and physical interactions.

\begin{figure}[h]
	\centering
	\includegraphics[width=0.99\linewidth]{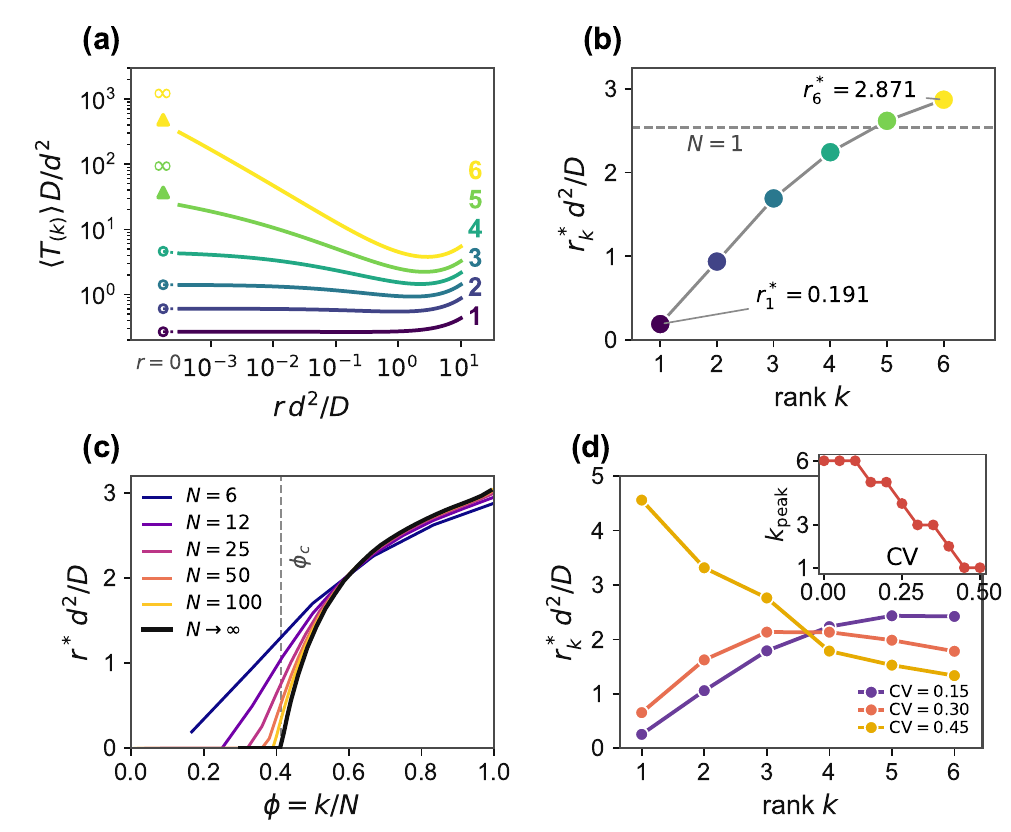}
	\caption{\textbf{Rank and initial geometry determine the optimal
			resetting rate of non-interacting Brownian searchers.}
		(a) Mean ordered first-passage time
		$\langle T_{(k)}\rangle D/d^2$ versus $rd^2/D$ for $N=6$
		searchers starting at the same distance $d$.
		(b) Optimal resetting rate $r_k^*d^2/D$ versus arrival rank.
		The dashed line marks the single-particle optimum
		$r_{\mathrm{1p}}^*d^2/D\simeq2.540$.
		(c) Optimal rate versus rank fraction $\phi=k/N$ for increasing
		$N$, showing convergence toward the large-$N$ quantile limit and the
		critical fraction $\phi_c\simeq0.412$.
		(d) Effect of initial-distance heterogeneity on the optimal-rate profile
		for a symmetric hexagon-like family at fixed mean distance $\bar d$.
		Increasing $\mathrm{CV}=\sigma_d/\bar d$ shifts the maximum from late
		toward earlier arrival ranks. Inset: peak rank $k_{\mathrm{peak}}$ versus
		$\mathrm{CV}$ for the same family.
	}
	\label{fig:homogeneous}
\end{figure}

Figure~\ref{fig:homogeneous}a displays the mean ordered first-passage times for
$N=6$. The sensitivity to stochastic resetting changes markedly across arrival
ranks: the interior minimum is shallow for the fastest arrival ($k=1$) but
deepens progressively for later arrivals. Consequently, the optimal resetting
rate $r_k^*$ increases monotonically with rank $k$ (Fig.~\ref{fig:homogeneous}b),
shifting from $r_1^*d^2/D \simeq 0.19$ to $r_6^*d^2/D \simeq 2.87$. Different
completion criteria thus select distinct optimal resetting rates, even in
homogeneous, non-interacting populations.

For comparison, a single Brownian searcher exhibits the well-known optimum
$r_{\mathrm{1p}}^*d^2/D \simeq 2.540$ (see Ref.\,\cite{EvansMajumdar2011}). The ordered-passage
optimum $r_k^*$ is therefore not governed by the single-particle benchmark:
it falls well below it for early arrivals and surpasses it for the final arrival.

This rank-dependent sensitivity plays a critical role for the latest arrivals.
Without resetting, the long-time survival probability of a one-dimensional
Brownian searcher decays as $Q_0(t) \sim t^{-1/2}$, yielding
\begin{equation}
	\Pr[T_{(k)}>t]\sim t^{-(N-k+1)/2}.
\end{equation}
For $N=6$, the mean arrival times for ranks $k=5$ and $k=6$ diverge at $r=0$.
Any finite resetting rate truncates these heavy-tailed excursions, rendering
the mean passage times finite. For earlier ranks, the relative depth of the resetting minimum increases
strongly with $k$ (Fig.~S1).

As the population size $N$ increases at fixed rank fraction $\phi=k/N$,
the finite-$N$ profiles converge toward the large-$N$ quantile limit
(Fig.~\ref{fig:homogeneous}c). In the $N\to\infty$ limit, the optimal
resetting rate vanishes below the critical fraction $\phi_c\simeq0.412$,
whereas a finite nonzero optimum persists above $\phi_c$, recovering the
quantile threshold identified by Belan~\cite{Belan2020}. The discrete
finite-$N$ sequence connects smoothly to this asymptotic limit. Notably, the
shallow finite optimum observed for $k=1$ at $N=6$ is consistent with the
critical-population behavior reported for independently reset searchers~\cite{BiroliMajumdarSchehr2023}.

We next examine spatial heterogeneity in non-interacting, independently reset
populations. Varying the spread of starting distances $\mathrm{CV} =\sigma_d/\bar{d}$ within a symmetric
hexagonal configuration at fixed mean distance $\bar{d}$ dramatically reshapes
the sequence $r_k^*$ (Fig.~\ref{fig:homogeneous}d). Increasing spatial dispersion shifts the maximum of $r_k^*$ from the final arrival ($k=N$), through intermediate ranks, to the earliest arrival ($k=1$).

Crucially, arrival rank does not map deterministically to a specific particle or
initial position. Instead, each starting distance contributes probabilistically
to each order statistic. Nearer searchers contribute preferentially to early arrivals, while more distant searchers contribute more strongly to late arrivals, with substantial statistical overlap. This rank
mixing (see SM Fig.~S2) alters the weighted superposition of distance-conditioned
first-passage distributions entering each $T_{(k)}$, thereby reorganizing the
sequence $r_k^*$.

For the two geometries studied experimentally, $\mathrm{CV} =0.589$ ($L = 3\,\mu\mathrm{m}$)
and $\mathrm{CV} = 0.442$ ($L = 4\,\mu\mathrm{m}$). Within the independently reset hexagonal family shown in Fig.~\ref{fig:homogeneous}d, both geometries lie beyond the crossover at which the maximum of $r_k^*$ occurs at $k_{\mathrm{peak}} = 1$. Geometric heterogeneity alone thus reorganizes the monotonic sequence seen in homogeneous systems, though it does not fully explain the intermediate-rank maxima observed in the physical colloidal experiments.

Although the coefficient of variation serves as a concise descriptor for symmetric
configurations, it is not universal. Distributions with similar $\mathrm{CV}$
values show comparable shifts in $k_{\mathrm{peak}}$, whereas asymmetric distributions
featuring isolated near or far outliers produce distinct rank profiles for identical
mean and variance (Fig.~S3).

Finally, the choice of resetting protocol introduces implicit statistical
correlations. Under local resetting, searchers reset independently, whereas under
global resetting, a shared Poisson clock resets all particles simultaneously. Even
in the absence of physical interactions, global resetting correlates individual
passage times. While $r_k^*$ still increases with rank for homogeneous searchers
under global resetting, the absolute optimal values shift quantitatively (Fig.~S4).
Benchmark comparisons for physical systems must therefore employ protocol-matched
non-interacting controls.

The non-interacting reference thus separates three baseline effects present
before physical interactions are introduced: arrival rank controls the
reset-rate dependence of $\langle T_{(k)}\rangle$, spatial geometry controls
the structure of the optimal-rate sequence, and the reset protocol determines
whether otherwise non-interacting searchers acquire additional temporal
correlations. 

% ---------------------------------------------------------------------------
% PHYSICAL SYSTEMS
% ---------------------------------------------------------------------------

\subsection{Physical systems beyond the independent reference}
\label{subsec:interacting_systems}

\begin{figure}[t]
	\centering
	\includegraphics[width=0.97\linewidth]{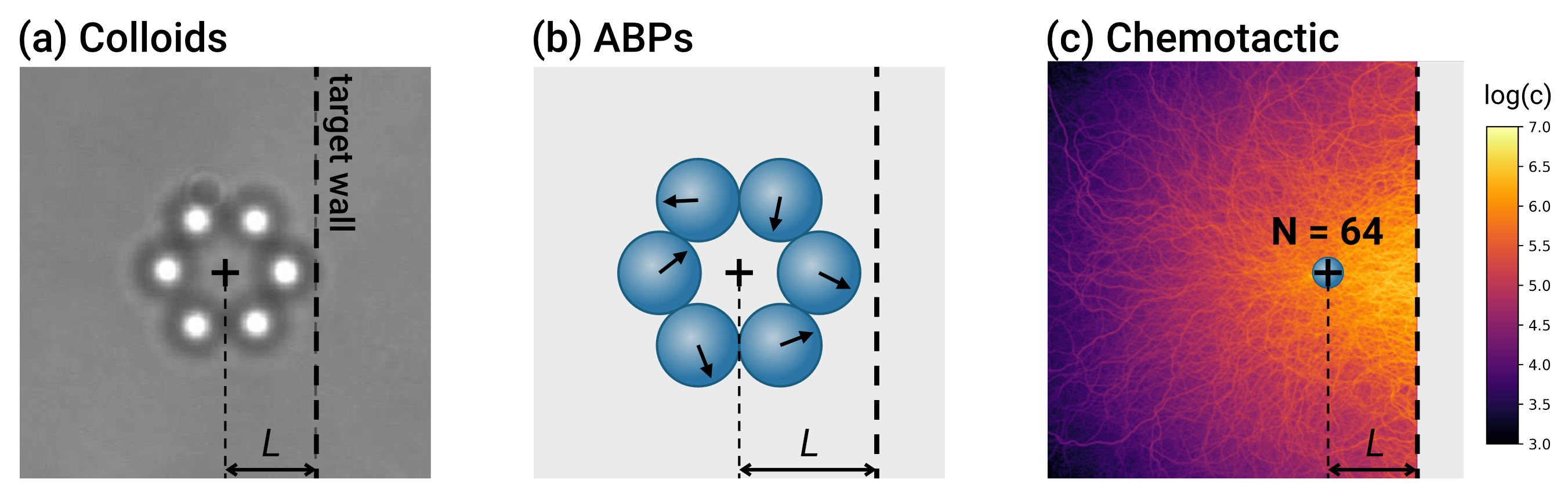}
	\caption{\textbf{Soft and active matter realizations of multiparticle
			search with resetting.}
		(a) Experimental setup of $N=6$ Brownian silica colloids subjected to
		global resetting via holographic optical tweezers. Particles interact
		through steric repulsion and hydrodynamic coupling.
		(b) Simulation of $N=6$ active Brownian particles (ABPs) with
		excluded-volume interactions under local resetting. Resetting relocates a
		particle to an available home site and randomizes its propulsion direction.
		(c) Simulation of $N=64$ autochemotactic active particles coupled through a
		self-generated chemoattractant field (heat map). Particles reset locally to
		the origin, leaving the accumulated chemical field intact to preserve
		environmental memory across resets.
		Detailed parameters and protocols are provided in
		Table~\ref{Table:Parameters} and Sec.~\ref{sec:Methods}.}
	\label{fig:systems}
\end{figure}

To evaluate how physical interactions and protocol constraints reshape the
non-interacting baseline, we examine three soft and active matter realizations
under stochastic resetting (Fig.~\ref{fig:systems}). In each system, $T_i$
denotes the first target encounter of labeled particle $i$, and the set
$\{T_i\}$ is ordered to construct the rank statistics $T_{(k)}$. Unlike the
independent baseline, physical searchers generally exhibit correlated passage
times, and dynamics following an early arrival can influence subsequent
arrivals. Only the first target encounter of each particle contributes to the
ordered arrival sequence.

The first realization is an optical-tweezer experiment with $N=6$ Brownian
silica colloids, utilizing the many-body resetting platform described in
Ref.~\cite{VatashRoichman2025}. Searchers are initialized at the vertices of a
hexagonal trap array before being released to diffuse toward a virtual target
line. Global reset events, driven by a shared Poisson process, return all
particles simultaneously to their initial positions. Because the physical
optical transport duration is subtracted from the first-passage clock, the
dynamics implement the instantaneous teleportation limit with zero return-time
overhead. Interparticle interactions arise from near-contact steric forces and
partially screened hydrodynamic coupling~\cite{HarelYael14,VatashRoichman2025}.

The second realization models $N=6$ interacting active Brownian particles
(ABPs). Each particle undergoes persistent self-propulsion, translational and
rotational diffusion, and short-range steric repulsion via a
Weeks--Chandler--Andersen (WCA) potential. Local reset events are generated by a single system-wide Poisson clock with rate $r_{\mathrm{sys}}$. At each event, one particle is chosen uniformly at random and returned to an unoccupied site in the six-site home configuration, with its orientation randomized upon return. The reset clock is suspended during the finite return duration $t_r = 4\,\mathrm{s}$; after return, a new exponential waiting time is drawn. Figures report the nominal per-particle rate $r=r_{\mathrm{sys}}/N$. This finite return time introduces an explicit protocol
overhead to the search duration. Removing $t_r$ in a numerical teleportation
control isolates the contribution of the finite return duration from the
free-search dynamics.

The third realization comprises $N=64$ autochemotactic active particles
coupled directly and through a self-generated chemoattractant field ~\cite{Rudyak2025,RudyakRoichman2026}.
Resetting is local: an individual particle is returned to the origin with a
randomized orientation, while all other searchers and the accumulated chemical
field remain unchanged. Local resetting therefore constitutes only a partial
reset of the many-body state. In the chemotactic system, this non-renewal has
an additional physical form: previous trajectories are stored in the
persistent chemical field and continue to influence later search dynamics
over the field-decay timescale.

\begin{table}[h]
	\centering
	\begin{tabular}{cccc}
		\hline\hline
		Parameter & Colloids & ABP & Chemotactic \\
		\hline
		$R$ ($\mu\mathrm{m}$) & 0.75 & 10 & 0.05 \\
		$L$ ($\mu\mathrm{m}$) & 3, 4 & 20.5, 30 & 2, 3, 5 \\
		$a_0$ ($\mu\mathrm{m}$) & 2.5 & 20 & 0 \\
		$D_T$ ($\mu\mathrm{m}^2/\mathrm{s}$) & 0.16 & 0.22 & 0 \\
		$D_R$ ($\mathrm{s}^{-1}$) & \text{---} & 0.16 & 0.16 \\
		$v_0$ ($\mu\mathrm{m}/\mathrm{s}$) & \text{---} & 12 & 0.16 \\
		$D_{\mathrm{eff}}$ ($\mu\mathrm{m}^2/\mathrm{s}$) & 0.16 & 450 & 0.08 \\
		$\tau_D$ ($\mathrm{s}$) & 14.1 & 0.23 & 12.5 \\
		$\tau_R$ ($\mathrm{s}$) & \text{---} & 6.25 & 6.25 \\
		\hline\hline
	\end{tabular}
	\caption{Representative physical parameters for the three search systems.
		Here $a_0$, $R$, and $L$ denote the initial interparticle spacing, particle
		radius, and target distance, respectively.
		We define the characteristic diffusive timescale as \(\tau_D=\frac{L^2}{4D_{\mathrm{eff}}}\), where \( D_{\mathrm{eff}}=D_T+\frac{v_0^2}{2D_R}\) is the long-time effective diffusion coefficient of the active Brownian particle (for a brownian particle \(D_{\mathrm{eff}}=D_T\)). The rotational timescale is \(\tau_R=1/D_R\). The tabulated values of $\tau_D$ use the shortest listed target distance $L$ for each system.
	}
	\label{Table:Parameters}
\end{table}

Because these physical systems incorporate distinct coupling mechanisms and
resetting constraints, they are not expected to follow a universal optimal-rate
curve. Instead, each is compared with an appropriate non-interacting reference
that accounts for its geometry, single-particle dynamics, and resetting
protocol as closely as possible. For the colloidal experiment, the reference
retains the heterogeneous hexagonal geometry and shared global reset clock.
For the ABPs, the non-interacting control retains the active dynamics,
finite-return protocol, and occupancy-dependent assignment of available home
sites while removing the WCA interaction. Because site assignment remains
occupancy dependent, this control is non-interacting in the sense of having no
direct interparticle forces, but is not strictly statistically independent.
For the autochemotactic system, we use a confined non-interacting active
baseline with the same propulsion, rotational diffusion, target geometry,
stochastic resetting, and boundary rules, but without direct repulsion,
chemical coupling, or target-reinforcement dynamics. The latter therefore
serves as an active baseline for the full chemotactic system rather than as a
one-parameter control of the chemical interaction.

% ---------------------------------------------------------------------------
% FIGURE 4 AND MECHANISTIC COMPARISON
% ---------------------------------------------------------------------------

\subsection{Physical search dynamics reshape rank-dependent optimal resetting}
\label{sec:interactions}

\begin{figure*}[t]
	\centering
	\includegraphics[width=\linewidth]{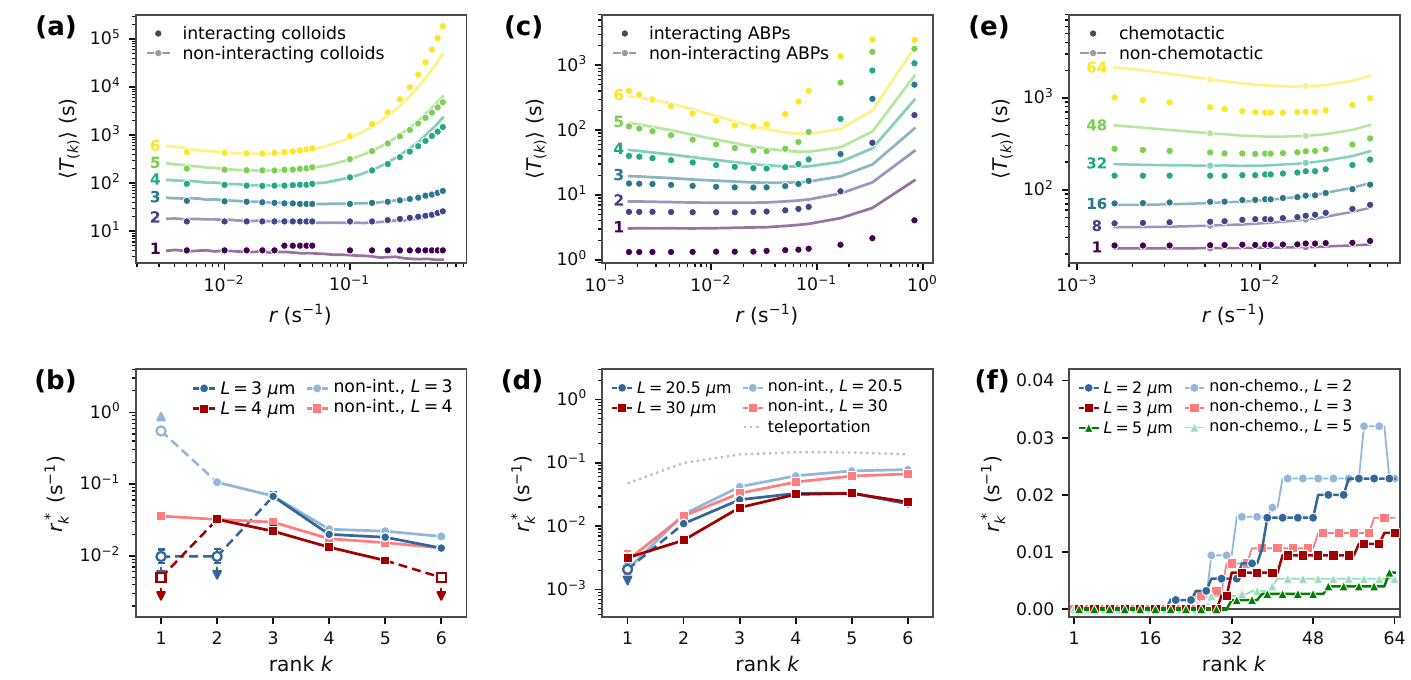}
	\caption{\textbf{Physical search dynamics reshape rank-dependent optimal
			resetting.}
		Top row: mean ordered first-passage times
		$\langle T_{(k)}\rangle$ versus resetting rate for the physical search
		systems (symbols) and their corresponding non-interacting references
		(solid lines), using $L=3\,\mu\mathrm{m}$ in panel (a),
		$L=20.5\,\mu\mathrm{m}$ in panel (c), and $L=5\,\mu\mathrm{m}$ in
		panel (e). Bottom row: optimal resetting rates $r_k^*$ extracted from
		these curves.
		(a,b) Brownian colloids under global resetting, compared with a
		non-interacting colloidal control using the same geometry
		and global-reset protocol. 
		(c,d) Interacting active Brownian particles under local resetting.
		Excluded-volume interactions and finite return overhead alter the
		optimal-rate sequence relative to the non-interacting control. The gray
		dashed curve labelled ``teleportation'' in panel (d) is the zero-overhead
		non-interacting control at $L=20.5\,\mu\mathrm{m}$ and isolates the
		contribution of the finite return duration.
		(e,f) Full autochemotactic system ($N=64$) compared with a confined
		non-interacting active baseline. A resolved finite-rate optimum appears
		only above a rank-dependent onset. For the colloidal and ABP systems, arrows
		with dashed connecting segments denote bounds where interior minima cannot
		be robustly resolved. For the chemotactic system, zero-valued points indicate ranks
		for which no beneficial finite-rate optimum is resolved on the sampled
		rate grid.}
	\label{fig:MKPTandR}
\end{figure*}

Across all three physical realizations, a common trend persists: early ordered
arrivals exhibit shallow or unresolved minima, whereas intermediate and late
arrivals develop increasingly robust finite-rate optima
(Fig.~\ref{fig:MKPTandR}a,c,e). The rank-dependent sensitivity established
by the non-interacting reference therefore remains central even when additional
physical couplings substantially alter the underlying search dynamics. These
additional ingredients are reflected in how the sequence $r_k^*$ departs from
the corresponding baseline  (Fig.\,\ref{fig:MKPTandR}b,d,f).

\subsubsection{Brownian colloids: interaction-induced probability redistribution}

For the Brownian colloids, the largest departures from the non-interacting
global-reset reference occur at early and intermediate ranks
(Fig.~\ref{fig:MKPTandR}a,b). The experiment develops an intermediate-rank
maximum in the optimal-rate sequence, with
$k_{\mathrm{peak}}=3$ for $L=3\,\mu\mathrm{m}$ and
$k_{\mathrm{peak}}=2$ for $L=4\,\mu\mathrm{m}$. This behavior goes beyond the
substantial reorganization already produced by the heterogeneous reset
geometry in the independent Brownian reference.

Previous measurements on the same optical-tweezer platform demonstrate that
interparticle interactions modify the positional statistics of the searchers
during resetting dynamics~\cite{VatashRoichman2025}. Hard-core steric
collisions redistribute probability toward larger displacements, whereas
hydrodynamic coupling partially counteracts this spatial spreading. Such
redistribution changes the trajectories contributing to the ordered arrivals
and provides a physical basis for the observed departure from the
non-interacting reference.

At later ranks, several distinct particles must explore the target region
within the same realization, so steric competition can further modify the
completion dynamics. We do not attempt to assign a unique microscopic
mechanism to each rank. Rather, the experiment demonstrates that
interaction-induced dynamics reorganize the rank-resolved optimum beyond the
effects generated by spatial heterogeneity and the global resetting protocol
alone.

\subsubsection{Active Brownian particles: steric interactions and finite return cost}

Interacting ABPs exhibit a distinct departure from their WCA-off reference
(Fig.~\ref{fig:MKPTandR}c,d). At low resetting rates, the interacting
ordered passage times are reduced relative to the control, particularly for
early arrivals. This behavior is consistent with excluded-volume interactions
favoring outward escape from the compact reset configuration, which can
enhance target-reaching trajectories for particles on the target-facing side
of the cluster.

At higher resetting rates, frequent local resets repeatedly repopulate the
compact home configuration, increasing the importance of excluded-volume
constraints. In addition, each physical return requires a finite duration
$t_r=4\,\mathrm{s}$, imposing a cumulative dead-time overhead that grows with
reset frequency. Correspondingly, the optimal resetting rates shift downward
relative to the non-interacting control and form a broad maximum at late
ranks.

The teleportation control isolates the contribution of the finite return
duration (Fig.~\ref{fig:MKPTandR}d). Eliminating this duration reduces the
high-rate penalty, showing that part of the shift in $r_k^*$ arises from the
physical implementation of resetting rather than from the free-search
dynamics alone. The ABP system therefore illustrates that finite return cost
can become an integral component of multiparticle resetting optimization.

\subsubsection{Autochemotactic particles: environmental memory across partial resets}

The autochemotactic system exhibits a qualitatively different departure from
its active baseline. We compare the full $N=64$ interacting chemotactic
dynamics with a confined non-interacting active reference sharing the same
propulsion speed $v_0$, rotational diffusivity $D_R$, domain size
($32\times32\,\mu\mathrm{m}$), target geometry, stochastic resetting, and
boundary rules (Fig.~\ref{fig:MKPTandR}e,f). Because direct repulsion and
target-reinforcement dynamics are also absent from this reference, the
comparison characterizes the full collective chemotactic system relative to
non-communicating active searchers rather than isolating chemical coupling
alone.

The resulting kinetics differ non-monotonically across arrival ranks
(Fig.~\ref{fig:MKPTandR}e). Chemotaxis delays the first arrival, and the
delay persists to an intermediate rank -- $k=23$, $19$ and $16$ for $L=2$, $3$ and $5\,\mu\mathrm{m}$ -- where both non-chemotactic and chemotactic systems complete the task in the same time. Before any particle reaches the target the chemoattractant field carries no information about its location, yet each searcher already lays a trail: the rotational response aligns the intrinsic direction along existing trails, while the translational response draws particles back into
them\cite{RudyakRoichman2026}. The population therefore re-runs a branching network of channels laid down blind instead of exploring fresh territory, which raises the first-passage time relative to non-chemotactic active searchers, as shown in detail for a single searcher\cite{Rudyak2025}. 
The first target encounter changes this. On contact the intrinsic direction is reversed, $\mathbf{n}_i\rightarrow-\mathbf{n}_i$, and secretion is transiently amplified by $\beta_{\mathrm{mult}}=100$, so the returning particle retraces its approach path while depositing a strongly enhanced trail, converting one blind branch into a directed channel linking the target back to the origin. Because local resetting leaves the field intact, this channel survives every reset and is reinforced by every
subsequent arrival. Arrival rank therefore acts as a clock for channel formation: the two systems break even at the ranks quoted above, earlier for more distant targets, and higher ranks are reached progressively faster in the chemotactic system. Resetting enters twice with opposite signs, delaying the initial channel formation but, once strong channels exist, helping to avoid long searches in channels that do not lead to the target. The two mechanisms give the chemotactic and non-chemotactic systems markedly different sensitivities of $\langle T_{(k)}\rangle$ to $r$, yet the extracted optimal rates $r_k^{*}$ remain close in both (Fig.~\ref{fig:MKPTandR}f).

The resulting kinetics differ substantially. At $L=3\,\mu\mathrm{m}$, for
example, the full chemotactic system reduces the final-arrival time at
$r=0$ from approximately $2769\,\mathrm{s}$ in the active baseline to
approximately $1258\,\mathrm{s}$, while the resolved final-arrival optimum
shifts from $r_N^*\approx0.016\,\mathrm{s}^{-1}$ to
$0.013\,\mathrm{s}^{-1}$ (Fig.~S5).

Chemical coupling also shifts the rank at which a finite-rate optimum first
becomes resolvable, but not in the direction of earlier ranks. The onset comparison uses a separate calculation on the rate grid common to both datasets: a finite-rate benefit is accepted only when the discrete minimum improves on the $r=0$ value by more than $1\%$. Neither the bootstrap nor the local quadratic fit used to extract the plotted optimal rates enters this calculation. This common criterion gives onset ranks
$k=27$, $27$ and $29$ in the active baseline for $L=2$, $3$ and
$5\,\mu\mathrm{m}$, against $k=27$, $32$ and $35$ in the chemotactic system:
the onset is unchanged at the shortest target distance and delayed by five and
six ranks at the two larger ones. The chemical field sustains the unreset
search for longer before restarting becomes worthwhile. The staircase structure in
Fig.~\ref{fig:MKPTandR}(f) reflects the discrete resetting-rate grid and the
finite resolution of the optimization procedure rather than a sequence of
discrete physical transitions.

In summary, the optimal-rate sequence $\{r_k^*\}$ serves as a sensitive, rank-resolved probe of physical search dynamics. Because spatial geometry and reset protocols generate substantial structure in $r_k^*$ even without particle interactions, physical coupling must be interpreted through deviations from appropriate non-interacting reference baselines that account
for geometry and resetting protocol.

% ---------------------------------------------------------------------------
% CONCLUSION
% ---------------------------------------------------------------------------

\section{Conclusions}

We have shown that stochastic-resetting optimization in a multiparticle
search is intrinsically rank dependent. A population is not characterized by
a single optimal resetting rate, but by a sequence $\{r_k^*\}$ determined by
the required completion rank. For non-interacting Brownian searchers,
$r_k^*$ increases strongly with rank for homogeneous initial conditions and
approaches the known large-$N$ quantile limit. For the latest Brownian
arrivals, resetting additionally regularizes mean passage times that diverge
in its absence. Spatial heterogeneity further shows that even without
interactions the optimal-rate sequence can become non-monotonic and shift its
maximum toward early ranks.

The physical systems considered here demonstrate how this baseline is
reshaped by additional search dynamics. Colloidal interactions modify the
optimal-rate profile beyond the substantial effect already produced by the
heterogeneous reset geometry. For active Brownian particles, excluded-volume
interactions and a finite return duration alter the optimum in distinct ways,
with a teleportation control exposing the contribution of the finite return
time. In the autochemotactic system, local resetting leaves both the rest of
the population and a self-generated chemical field intact, allowing previous
trajectories and successful arrivals to modify the environment encountered by
later searchers.

The broader implication is that optimization of collective search should be
formulated in terms of the required completion rank, or equivalently the
required fraction of successful searchers in large populations. Moreover, the
shape of $\{r_k^*\}$ cannot by itself be assigned to collective interactions:
ordered statistics, reset geometry, and reset protocol already generate
substantial rank dependence. An appropriate non-interacting baseline is
therefore essential for interpreting how physical coupling and environmental
memory reshape multiparticle search.

% ---------------------------------------------------------------------------
% MODELS AND METHODS
% ---------------------------------------------------------------------------

\section{Models and methods}
\label{sec:Methods}

We investigate three representative physical search platforms under stochastic resetting: an experimental suspension of Brownian colloids, a simulation of interacting active Brownian particles (ABPs), and a simulation of autochemotactic active particles. For each realization, the first target encounter of labeled particle $i$ defines its individual first-passage time $T_i$. The set of first-passage times $\{T_i\}_{i=1}^N$ is ordered to construct the rank-resolved arrival sequence $T_{(1)} \leq \dots \leq T_{(N)}$. Secondary target encounters by an already arrived particle do not generate additional ordered arrivals. System-specific post-arrival dynamics are detailed below.

\subsection{Brownian colloids under optical-tweezer resetting}
\label{subsec:Colloids}

The experimental system comprises six charge-stabilized silica colloids of diameter $d_p = 1.5 \pm 0.08\,\mu\mathrm{m}$ dispersed in double-distilled water. The particles sediment near the lower glass surface of the sample cell, where the measured single-particle diffusion coefficient is $D = 0.164 \pm 0.04\,\mu\mathrm{m}^2\,\mathrm{s}^{-1}$. Particle trajectories are recorded at 30 frames per second using a Grasshopper3 camera (Point Grey) and tracked via standard video-microscopy algorithms~\cite{CROCKER1996VideoAnalysis}.

Interparticle forces combine near-hard-core repulsion and partially screened hydrodynamic interactions~\cite{HarelYael14,VatashRoichman2025}. Previous measurements on this platform demonstrated that steric collisions redistribute probability toward larger displacements, whereas hydrodynamic coupling partially opposes this spatial broadening~\cite{VatashRoichman2025}.

Six holographic optical traps are configured at the vertices of a regular hexagon centered at the origin~\cite{Dufresne_Grier01}, with a nearest-neighbor trap spacing of $a_0 = 2.5\,\mu\mathrm{m}$. Each realization begins with one particle confined per trap. The traps are extinguished simultaneously, allowing the particles to diffuse freely.

A planar virtual target line is positioned at $x = L$, with experiments conducted at $L = 3\,\mu\mathrm{m}$ and $L = 4\,\mu\mathrm{m}$. A first-passage event occurs when a \emph{particle center} crosses the target line:
\begin{equation}
	T_i = \inf \{t > 0 : x_i(t) \ge L\}.
\end{equation}
All distances reported represent center-to-target-line separations.

For a hexagonal array oriented with an axis toward the target, initial particle distances are given by $d_i = L - a_0 \cos\theta_i$, where $\theta_i$ is the angular position of trap $i$. For $L = 3\,\mu\mathrm{m}$, the distance set is $\{d_i\} = \{0.5, 1.75, 1.75, 4.25, 4.25, 5.5\}\,\mu\mathrm{m}$, yielding a coefficient of variation $\mathrm{CV} = 0.589$. For $L = 4\,\mu\mathrm{m}$, $\{d_i\} = \{1.5, 2.75, 2.75, 5.25, 5.25, 6.5\}\,\mu\mathrm{m}$, yielding $\mathrm{CV} = 0.442$.

Stochastic resetting is controlled by an automated holographic protocol. Waiting times between resets are drawn from the exponential distribution $p(t) = r e^{-rt}$, corresponding to a Poisson reset process with system-wide rate $r$ and a single shared Poisson clock. At each reset event, all six particles return to their initial trap positions.

In the first-passage analysis, the optical return stage is treated as instantaneous: return trajectories and transport overhead are excluded from the search duration. Free search resumes once all particles are restored to their initial positions, matching global teleportation resetting with zero return-time cost. Tested resetting rates span $0.005\,\mathrm{s}^{-1} \leq r \leq 0.6\,\mathrm{s}^{-1}$. Realizations run until every distinct particle crosses the target line, yielding a complete rank sequence $T_{(1)}, \dots, T_{(6)}$ per run. The procedure is repeated $10^4$ times for each
resetting rate, and ensemble averages yield the mean ordered passage times
$\langle T_{(k)}\rangle$.

\subsection{Interacting active Brownian particles}
\label{subsec:ABPs}

The second system models $N = 6$ interacting active Brownian particles in two dimensions. During free motion, the position $\mathbf{r}_i$ of particle $i$ evolves via
\begin{equation}
	d\bm{r}_i= \left( v_0\hat{\bm{n}}_i + \mu\bm{F}_i \right)dt
	+
	\sqrt{2D_T}\,d\bm{W}_i,
\end{equation}
where $v_0$ is the self-propulsion speed, $\mu$ is the translational mobility, and $\mathbf{F}_i$ is the total conservative interaction force. The propulsion orientation is $\hat{\mathbf{n}}_i = (\cos\theta_i, \sin\theta_i)$, with orientation dynamics governed by
\begin{equation}
	d\theta_i
	=
	\sqrt{2D_R}\,dW_i^{(R)},
\end{equation}
where $D_R$ is the rotational diffusion coefficient. 

Here \(d\bm{W}_i\) and \(dW_i^{(R)}\) are independent standard
translational and rotational Wiener increments, respectively, satisfying
\begin{align}
	\left\langle dW_{i,\alpha} \right\rangle &= 0, &
	\left\langle dW_{i,\alpha}dW_{j,\beta} \right\rangle
	&= \delta_{ij}\delta_{\alpha\beta}\,dt, \\
	\left\langle dW_i^{(R)} \right\rangle &= 0, &
	\left\langle dW_i^{(R)}dW_j^{(R)} \right\rangle
	&= \delta_{ij}\,dt,
\end{align}

with the translational and rotational Wiener processes mutually independent.

Particles interact through a short-ranged Weeks--Chandler--Andersen (WCA) pair potential~\cite{WCA}, with \(T=298\,\mathrm{K}\), \(\epsilon=k_{\mathrm B}T\), \(\sigma=2R\), and \(\mu=D_T/(k_{\mathrm B}T)\). The simulations are integrated using a timestep of \(\Delta t=0.002\,\mathrm{s}\). Hydrodynamic interactions are neglected.

Initially, the six particles occupy the vertices of a regular hexagon centered at the origin with nearest-neighbor spacing $a_0 = 20\,\mu\mathrm{m}$. The home positions are
\begin{equation}
	\mathbf{r}_m^{\mathrm{home}} = a_0 \left( \cos\frac{2\pi m}{6}, \sin\frac{2\pi m}{6} \right), \qquad m = 0, \dots, 5.
\end{equation}
A target boundary is located at distance $L$ from the origin, with primary simulations conducted at $L = 20.5\,\mu\mathrm{m}$, and \(L = 30\,\mu\mathrm{m}\).

Two return mechanisms are implemented. Upon reaching the target for the first time, a particle records its first-passage time and is restored to the nearest unoccupied home site. The arrived particle remains in the domain and continues interacting, but subsequent target contacts are ignored.

Stochastic resetting is driven by a single system-wide Poisson clock with rate $r_{\mathrm{sys}}$, in the range \(0.01\,\mathrm{s}^{-1}\leq r_{\mathrm{sys}}\leq 5\,\mathrm{s}^{-1}\). At each reset event, one particle is selected uniformly at random and returned to the nearest unoccupied home position. Upon completion of the return, its propulsion orientation is randomized. The reset clock is suspended during the return duration $t_r=4\,\mathrm{s}$, and a new exponential waiting time is drawn after the particle arrives at its home position. 

To obtain robust statistics, we performed $10^5$ independent simulations for each resetting rate.
Because of computational constraints, each realization was terminated after a
finite maximum duration of \(5000\,\mathrm{s}\). Within this observation window, all six particles
reached the target in approximately $95\%$ of the realizations, allowing us to
record the complete sequence of ordered first-passage times. The sampling uncertainty in the estimated mean ordered first-passage time is quantified by its standard error.

Figures report the nominal per-particle resetting rate $r=r_{\mathrm{sys}}/N$.

Physical return requires a finite duration $t_r = 4\,\mathrm{s}$, which contributes directly to elapsed passage times. During return, self-propulsion and stochastic translational and rotational dynamics are suspended for the returning particle, while steric interactions remain active. For assigned home position $\mathbf{r}_i^{\mathrm{home}}$ and remaining return time $t_{i,\mathrm{rem}}$, the deterministic return velocity is
\begin{equation}
	\mathbf{v}_{i,\mathrm{ret}} = \frac{\mathbf{r}_i^{\mathrm{home}} - \mathbf{r}_i}{t_{i,\mathrm{rem}}},
\end{equation}
updated continuously to ensure arrival at $t_{i,\mathrm{rem}} = 0$.

A zero-overhead teleportation control ($t_r = 0$) utilizing the same reset events and home-assignment rules is evaluated to isolate the contribution of the finite return duration from the free-search dynamics.

\subsection{Autochemotactic active particles}
\label{subsec:Chemotactic}

The third system consists of $N = 64$ autochemotactic active particles in two dimensions, combining self-propulsion, steric repulsion, and coupling via a self-generated chemoattractant field~\cite{Rudyak2025,RudyakRoichman2026}. Each particle $i$ has position $\mathbf{r}_i(t)$ and intrinsic orientation $\mathbf{n}_i(t) = (\cos\phi_i, \sin\phi_i)$. Translational thermal diffusion is neglected ($D_T = 0$), and orientation undergoes rotational diffusion with rate $D_R$.

Simulations are conducted in a square domain of size $32 \times 32$ centered at the origin $\mathbf{0}$. Particles originate at the center of the domain ($\mathbf{r}_i(0) = \mathbf{0}$)
with independently randomized intrinsic orientations.

\subsubsection{Chemoattractant field and environmental memory}
Particles move at constant speed $v_0=1$ and deposit a scalar chemoattractant field $c(\mathbf{x},t)$, initialized at $c(\mathbf{x},0) = 0$. Each particle secretes a Gaussian chemical profile with baseline amplitude $\beta_i = 1/64$ and spatial width $R_c = 0.01$. Equations of motion are

\begin{equation}
	\begin{aligned}
		&\mathrm{d}\mathbf{r}_i = v_0 \hat{\mathbf{u}}_i(\mathbf{r}_i,t)\,\mathrm{d}t, \\
		&\mathbf{u}_i(\mathbf{r}_i,t) = \mathbf{n}_i + \mathbf{f}_i(\{\mathbf{r}\}) + \chi_T \frac{\nabla c(\mathbf{r}_i,t)}{c(\mathbf{r}_i,t) + c_0}, \\
		&\hat{\mathbf{u}}_i = \frac{\mathbf{u}_i}{|\mathbf{u}_i|}, \\
		&\mathrm{d}\phi_i = \chi_R \frac{|\nabla c(\mathbf{r}_i,t)| \sin[2\psi_i(t)]}{c(\mathbf{r}_i,t) + c_0}\,\mathrm{d}t + \sqrt{2D_R}\,\mathrm{d}W_i,
	\end{aligned}
\end{equation}

where $\mathbf{f}_i(\{\mathbf{r}\})$ is total particle interaction force acting on particle $i$, $\psi_i$ is the angle between $\mathbf{n}_i$ and the local chemical gradient $\nabla c$. Chemotactic sensitivities are set to $\chi_T = 0.008$ and $\chi_R = 0.2$, $c_0=0.001$ is the
sensitivity noise level.

The chemical field evolves according to
\begin{equation}
	\partial_t c(\mathbf{x},t) = -\frac{1}{\tau_c} c(\mathbf{x},t) + \sum_j \beta_j \mathcal{N}_{0,R_c}\left(|\mathbf{x} - \mathbf{r}_j(t)|\right),
	\label{eq:chemical_field}
\end{equation}
with decay time $\tau_c = 100$.

Crucially, the chemoattractant field is not reset when an individual particle returns to the origin. Trajectory history remains encoded in the environment, creating a persistent memory that influences subsequent search dynamics.

\subsubsection{Direct particle interactions}

Particles interact directly through a pairwise soft harmonic repulsion potential:
\begin{equation}
	U(r_{ij}) = \frac{U_0\sigma}{2} \left( 1 - \frac{r_{ij}}{\sigma} \right)^2 \quad \text{for } r_{ij} < \sigma,
\end{equation}
and $U(r_{ij}) = 0$ for $r_{ij} \ge \sigma$. The total direct force on particle $i$ is denoted $\mathbf{f}_i(\{\mathbf{r}\})$. Simulations set $U_0 = 10$ and $\sigma = 0.01$.

\subsubsection{Resetting and boundary rules}

Each particle undergoes local Poisson resetting to the origin with the same resetting rate $r$. Upon resetting, its position returns to $\mathbf{0}$ instantaneously and its orientation is randomized, while other particles and the chemical field remain undisturbed. Particles reaching non-target outer boundaries are likewise returned instantaneously to the origin with randomized orientation.

\subsubsection{Target interaction and ordered arrivals}

A planar target is placed at distance $L$ from the origin ($L = 2, 3, 5\,\mu\mathrm{m}$). The first target contact of particle $i$ records its first-passage time $T_i$, while the particle remains dynamically active. The target is absorbing only for first-passage bookkeeping. Upon contact, the particle reverses its intrinsic orientation ($\mathbf{n}_i \to -\mathbf{n}_i$), and transiently amplifies its secretion rate by $\beta_{\mathrm{mult}} = 100$, which relaxes back toward baseline over the timescale $\tau_\beta = \frac{2L}{v_0}$.
Arrived particles continue interacting with other particles and modifying the chemical field, though subsequent target contacts are excluded from higher-order rank statistics.

To reach enough statistics and reduce measure errors, for each distance to the target $L$ and each resetting rate $r$, we calculated multiple independent realizations, from 1024 (for higher $r$ values) to 6144 (for lower $r$ values). The ensemble averages yield the mean ordered passage times $\langle T_{(k)}\rangle$.

\subsubsection{Scaling of the dimensionless simulations}
Chemotactic simulations are produced in the dimensionless form, and the results are scaled to experimental units afterwards together with the input parameters. We used scaling factors of $1\,\mu\text{m}$ for spatial scale and 6.25\,s for temporal scale. It effectively translates the simulated system into $32\times32\,\mu\text{m}^2$ simulation box, and particle radius $R=0.05\,\mu\text{m}$, speed $v_0=0.16\,\mu\text{m}/\text{s}$, rotational diffusion $D_R=0.16\,\text{s}^{-1}$, and distance to the target $L$ varied from $2\,\mu\text{m}$ to $5\,\mu\text{m}$.

\subsection{Non-interacting reference systems}
\label{subsec:controls}

To separate baseline effects associated with arrival rank, spatial geometry, and resetting protocol from additional physical search dynamics, each physical realization is compared with an appropriate non-interacting reference:

\begin{itemize}
	\item \textbf{Colloids:} Non-interacting Brownian searchers sharing the experimental six-distance initial setup and global Poisson reset clock. The shared reset clock is therefore retained, so correlations generated by global resetting remain present in the reference.
	\item \textbf{ABPs:} A WCA-off control matching $v_0$, $D_T$, $D_R$, target distance, initial geometry, finite-return protocol, and occupancy-dependent assignment of available home sites. Because site assignment remains occupancy dependent, this reference contains no direct interparticle forces but is not strictly statistically independent.
	\item \textbf{Chemotactic particles:} A confined active baseline ($N = 64$) sharing the same propulsion, rotational diffusion, domain boundaries, reset kinetics, target geometry, and non-target boundary-return rules, but excluding direct repulsion, chemotactic coupling, chemical deposition, and target-reinforcement dynamics. This system therefore provides an active baseline for the full chemotactic dynamics rather than a one-parameter control of chemical coupling.
\end{itemize}
\subsection{Extraction of optimal resetting rates}
\label{subsec:rstar_extraction}

For each arrival rank $k$ the optimal resetting rate is
\begin{equation}
	r_k^* = \operatorname*{arg\,min}_{r \ge 0} \, \langle T_{(k)}(r) \rangle.
	\label{eq:rstar}
\end{equation}
For the exact Brownian curves of Fig.~\ref{fig:homogeneous} this is evaluated by
direct numerical minimization over $\log_{10} r$. For measured and simulated
data (Fig.~\ref{fig:MKPTandR}) a local quadratic $y = Ax^2+Bx+C$ is fitted in
$x = \log_{10} r$ around the discrete minimum, giving $r^* = 10^{-B/(2A)}$.

The fitting window and noise model differ by system. Colloids: seven points
centred on the minimum, weighted by $1/\sigma_i^2$ with
$\sigma_{\mathrm{eff}} = \sqrt{\mathrm{SEM}^2 + 1/12}$, the second term covering
integer-second rounding. ABPs: five points, with one noise scale from the
degree-of-freedom-corrected RMS residual of that fit. Chemotactic simulations,
which carry no reported uncertainties: one $\sigma$ per rank from pooled
residuals of sliding five-point quadratics along the whole curve. Where the
window would run off the sampled grid it is replaced rather than truncated --- by
the first seven points (colloids) or the last five (chemotactic), while for the
ABPs a window shorter than three points suppresses the fit and the discrete
minimum is reported.

Uncertainties come from a parametric bootstrap ($n_{\mathrm{boot}} = 4000$,
$y_i^{(b)} = y_i + \mathcal{N}(0,\sigma_i)$), refitting each draw and keeping the
vertex only if $A>0$ and it lies inside the window. The chemotactic fits add one
condition, that the value at the vertex fall below an independently perturbed
$r=0$ value, and re-locate the window on the minimum of each draw, which lets two
shallow competing basins exchange order; the colloid and ABP windows stay fixed
at the minimum of the unperturbed curve. Two acceptance criteria then apply in
sequence: the bootstrap distribution is used only if at least $20\%$ of draws
give interior minima, and the resulting median is reported as a finite $r_k^*$
(with the 16th and 84th percentiles as asymmetric bars) only if fewer than half
of all draws were rejected and the discrete minimum is not at either end of the
grid. Ranks failing either test are reported as bounds --- arrows with dashed
connecting segments for colloids and ABPs, and $r_k^*=0$ for the chemotactic
system, meaning no beneficial finite rate is resolved within the sampled range.

None of the three non-interacting controls is a direct minimization. The colloid
reference is a global-resetting Monte-Carlo calculation of the finite-$N$ order
statistics on the same rate grid, passed through the same seven-point weighted
bootstrap with $\sigma_i$ the Monte-Carlo standard error, and reported with the
symmetric half-width of the resulting interval. The ABP reference shares the
five-point window but is not bootstrapped; its uncertainty is the spread over
three simulation seeds. The confined non-chemotactic reference is not fitted at
all: it is the discrete grid minimum, kept only where it improves on
$\langle T_{(k)}\rangle(0)$ by more than $0.2\%$, giving a stepped optimal-rate profile. The onset-rank comparison instead applies the same discrete-grid rule with a $1\%$ improvement threshold to both the control and chemotactic datasets, separately from the extraction of the plotted optimal rates. In the $32\times32\,\mu\mathrm{m}^2$ domain, a finite-rate optimum is resolved at $L=5\,\mu\mathrm{m}$ for sufficiently high arrival ranks (Fig.~\ref{fig:MKPTandR}f).

Two conversions are applied after extraction, so plotted rates differ from those
in the source files: chemotactic rates are multiplied by
$\tau_R^{-1} = D_R = 0.16\,\mathrm{s}^{-1}$ (Eq.~\eqref{eq:rstar} is invariant
under this rescaling), and ABP rates, stored as system-wide values, are divided
by $N=6$, so every panel shows a per-particle rate. Finally, for the
$L=4\,\mu$m colloidal series the four highest rates ($r\ge0.4\,\mathrm{s}^{-1}$)
are excluded because, at these highest resetting rates and this target distance,
the finite experimental duration yielded too few target arrivals to obtain
statistically reliable first-passage estimates.

%%%END OF MAIN TEXT%%%

%The \balance command can be used to balance the columns on the final page if desired. It should be placed anywhere within the first column of the last page.

%If notes are included in your references you can change the title from 'References' to 'Notes and references' using the following command:
%\renewcommand\refname{Notes and references}

\providecommand*{\mcitethebibliography}{\thebibliography}
\csname @ifundefined\endcsname{endmcitethebibliography}
{\let\endmcitethebibliography\endthebibliography}{}

\end{document}

\preprint{APS/123-QED}

\title{Supplementary Information: Rank-dependent optimal resetting in multiparticle search}
%\thanks{A footnote to the article title}%

\author{Ron Vatash}
\affiliation{The Raymond and Beverley School of Chemistry, Tel Aviv University, Tel Aviv 6997801, Israel.}
\author{Eden Goldfarb}
\affiliation{The Raymond and Beverley School of Physics \& Astronomy, Tel Aviv University, Tel Aviv 6997801, Israel.}
\author{Vladimir Yu. Rudyak}
\affiliation{The Raymond and Beverley School of Physics \& Astronomy, Tel Aviv University, Tel Aviv 6997801, Israel.}
\author{Yael Roichman}
\affiliation{The Raymond and Beverley School of Chemistry, Tel Aviv University, Tel Aviv 6997801, Israel.}
\affiliation{The Raymond and Beverley School of Physics \& Astronomy, Tel Aviv University, Tel Aviv 6997801, Israel.}

\date{\today}

\maketitle

\section{Strength of the rank-dependent resetting optimum}

\begin{figure}[h]
	\centering
	\includegraphics[width=0.7\linewidth]{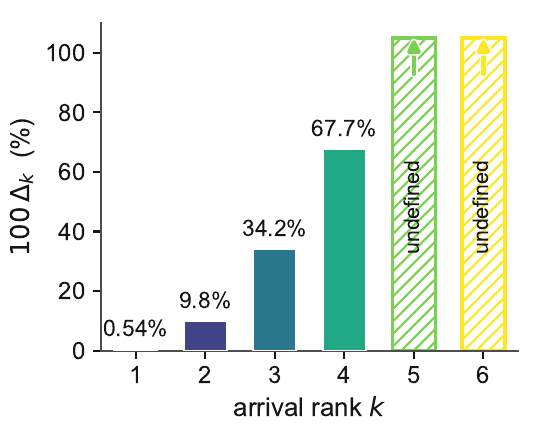}
	\caption{\textbf{Strength of the optimal resetting effect.}
		Relative reduction
		$\Delta_k=1-\langle T_{(k)}\rangle(r_k^*)/
		\langle T_{(k)}\rangle(0)$
		for $N=6$ identical non-interacting Brownian searchers.
		The improvement increases strongly with arrival rank.
		For $k=5$ and $6$, the zero-reset mean passage times diverge, so
		$\Delta_k$ is not defined; these ranks are indicated separately.}
	\label{fig:optimum_depth}
\end{figure}

The optimal resetting rate does not by itself indicate how strongly resetting
improves the passage time. To quantify the depth of the minimum, we define
\[
\Delta_k =
1-\frac{\langle T_{(k)}\rangle(r_k^*)}
{\langle T_{(k)}\rangle(0)} ,
\]
for ranks for which the mean passage time without resetting is finite.
As shown in Fig.~\ref{fig:optimum_depth}, the improvement is extremely weak
for the first arrival, with $\Delta_1\simeq0.54\%$, but increases rapidly with
rank, reaching approximately $9.8\%$, $34.2\%$, and $67.7\%$ for
$k=2$, $3$, and $4$, respectively.

For the last two arrivals the comparison with the unreset system becomes
singular. The long-time survival probability of a freely diffusing particle
decays as $Q_0(t)\sim t^{-1/2}$, so that
$\Pr[T_{(k)}>t]\sim t^{-(N-k+1)/2}$. Consequently, for $N=6$ the mean
fifth- and sixth-passage times diverge in the absence of resetting.
Any finite resetting rate suppresses these long excursions and renders the
corresponding mean passage times finite. Thus, while the first-arrival optimum
is mathematically finite but very shallow, resetting becomes progressively
more consequential for later ordered arrivals.
\section{How heterogeneous reset geometry reorganizes ordered arrivals}
\label{sec:SI_heterogeneity}

The main text shows that spatial heterogeneity of the reset positions can
qualitatively reorganize the sequence of optimal resetting rates
$\{r_k^*\}$ even for non-interacting searchers. Here we examine the
mechanism underlying this effect and clarify the extent to which it can be
summarized by the coefficient of variation
$\mathrm{CV}=\sigma_d/\bar d$.

\subsection{Rank mixing in a heterogeneous population}

Consider $N=6$ independently reset Brownian searchers with the same
diffusion coefficient but different initial distances $d_i$ from the
target. For identical starting positions, all particles are statistically
equivalent and the optimal resetting rate increases monotonically with
arrival rank. Distance heterogeneity removes this equivalence: particles
starting closer to the target contribute preferentially to early ordered
arrivals, whereas particles starting farther away contribute more strongly
to late arrivals.
\begin{figure}[h!]
	\centering
	\includegraphics[width=0.9\linewidth]{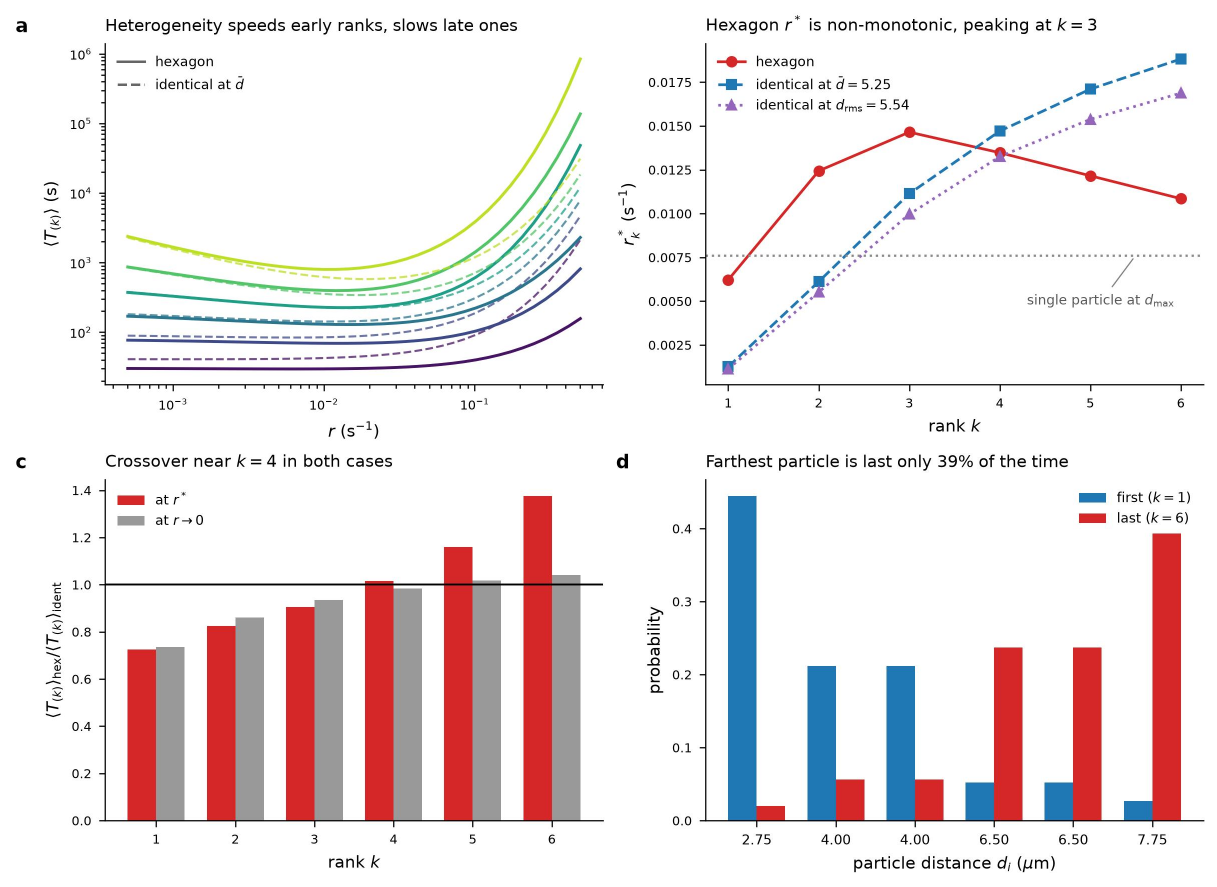}
	\caption{\textbf{Mechanism by which spatial heterogeneity reorganizes ordered first passage.}
		Results are shown for $N = 6$ independently reset Brownian searchers in a representative hexagonal geometry with starting distances
		$d_i = \{2.75, 4, 4, 6.5, 6.5, 7.75\}\,\mu\mathrm{m}$,
		$\bar{d} = 5.25\,\mu\mathrm{m}$, and $\mathrm{CV} \simeq 0.34$.
		\textbf{(a)}~Mean ordered passage times $\langle T_{(k)}\rangle$ versus resetting rate for the heterogeneous geometry (solid lines) and for six identical searchers starting at the same mean distance $\bar{d}$ (dashed lines). Heterogeneity preferentially shortens the early ordered passages and lengthens the late ones.
		\textbf{(b)}~Corresponding optimal resetting rates $r_k^*$. The heterogeneous geometry produces a non-monotonic rank dependence with a maximum at $k = 3$, in contrast to the monotonic increase for identical searchers. Identical populations initialized at $\bar{d}$ and at the root-mean-square distance $d_{\mathrm{rms}}$ are shown for comparison; the horizontal line marks the single-particle optimum for the largest starting distance $d_{\max}$.
		\textbf{(c)}~Ratio of the heterogeneous to identical mean ordered passage times as a function of rank, evaluated at optimal resetting (red) and in the zero-reset limit (gray). The crossover near $k = 4$ highlights the opposite effect of heterogeneity on early and late arrivals.
		\textbf{(d)}~Probability that each particle contributes the first ($k = 1$) or last ($k = 6$) arrival, plotted against its starting distance. Starting distance strongly biases arrival order but does not determine it uniquely: the nearest particle arrives first in only about $44\%$ of realizations, while the farthest particle arrives last in only about $39\%$.}
	\label{fig:heterogeneity_mechanism}
\end{figure}

Figure~\ref{fig:heterogeneity_mechanism} illustrates this effect for a
representative hexagonal geometry with
$\bar d=5.25\,\mu\mathrm{m}$ and $\mathrm{CV}\simeq0.34$.
Relative to six identical searchers starting at the same mean distance,
heterogeneity reduces the mean passage times of the early ranks while
increasing those of the later ranks
[Fig.~\ref{fig:heterogeneity_mechanism}(a)].
The corresponding sequence of optimal resetting rates is consequently
reorganized from a monotonic increase with rank to a non-monotonic profile,
with its maximum at an intermediate rank
[Fig.~\ref{fig:heterogeneity_mechanism}(b)].

Importantly, arrival rank is not uniquely determined by starting distance.
A particle beginning closest to the target need not arrive first, and the
farthest particle need not arrive last. As shown in
Fig.~\ref{fig:heterogeneity_mechanism}(d), the nearest particle contributes
the first arrival in only about $44\%$ of realizations, while the farthest
particle contributes the last arrival in only about $39\%$. Each ordered
passage time therefore samples a mixture of trajectories originating from
different reset distances, with the composition of this mixture changing
with rank.

This rank mixing provides the physical origin of the geometry dependence of
$r_k^*$. Resetting affects trajectories originating at different distances
differently, while each ordered passage samples a different weighted
combination of these distance-conditioned first-passage statistics.
Increasing the spread of the reset positions can therefore shift the maximum
of $r_k^*$ from the latest arrival through intermediate ranks and eventually
to the earliest arrival.

For the two reset geometries used in the colloidal experiment,
$\mathrm{CV}=0.589$ for $L=3\,\mu\mathrm{m}$ and
$\mathrm{CV}=0.442$ for $L=4\,\mu\mathrm{m}$. Within the independently
reset hexagonal reference considered in the main text, both lie beyond the
crossover to a maximum at the earliest rank, $k_{\rm peak}=1$.
Thus, spatial heterogeneity alone strongly reorganizes the homogeneous
rank dependence, but does not account for the intermediate-rank maxima
observed in the interacting colloidal experiment.

\section{Generality and limitations of the heterogeneity criterion}

\begin{figure}[h!]
	\centering
	\includegraphics[width=0.7\linewidth]{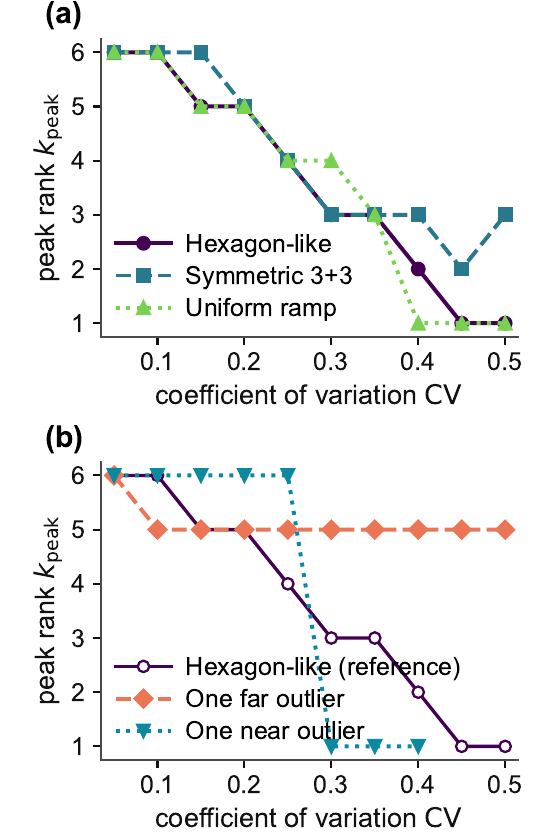}
	\caption{\textbf{Generality and limitations of the heterogeneity
			description.}
		(a) Rank $k_{\rm peak}$ at which the optimal resetting rate is maximal
		as a function of the coefficient of variation
		$\mathrm{CV}=\sigma_d/\bar d$ for three symmetric families of starting
		distances: the hexagon-like distribution, a symmetric $3+3$ bimodal
		distribution, and a uniform ramp. All show a similar migration of
		$k_{\rm peak}$ from late toward earlier ranks as heterogeneity increases.
		(b) The same analysis for asymmetric distance distributions containing a
		single far or near outlier. Despite comparable values of $\mathrm{CV}$,
		these configurations produce qualitatively different
		$k_{\rm peak}(\mathrm{CV})$. The hexagon-like result is shown for
		reference.}
	\label{fig:shape_dependence}
\end{figure}

The preceding analysis shows that distance heterogeneity reorganizes the
optimal-rate sequence by changing the mixture of starting distances
contributing to each arrival rank. We now ask whether the strength of this effect can be summarized by the coefficient of variation
$\mathrm{CV}=\sigma_d/\bar d$.

Figure~\ref{fig:shape_dependence}a compares three symmetric families of
starting distances at fixed mean distance: the hexagon-like family used in
the main text, a symmetric $3+3$ bimodal distribution, and a uniform ramp.
All three show the same qualitative progression: increasing heterogeneity
moves the maximum of $r_k^*$ from late toward progressively earlier arrival
ranks. The precise crossover values differ somewhat between distributions,
but the overall behavior is similar. Thus, for sufficiently symmetric
configurations, $\mathrm{CV}$ provides a useful compact measure of the
geometric heterogeneity relevant to the rank dependence of optimal
resetting.

This reduction to a single parameter fails for strongly asymmetric
configurations. In Fig.~\ref{fig:shape_dependence}b, introducing a single
far or near outlier produces markedly different $k_{\rm peak}$ even at
comparable values of $\mathrm{CV}$. A distant outlier preferentially
influences the late-ordered arrivals, whereas a nearby outlier strongly
biases the early ranks. Consequently, configurations with the same mean
distance and variance need not produce the same optimal-rate sequence.

The coefficient of variation should therefore be viewed as a convenient
descriptor for the symmetric family considered in the main text, rather
than as a universal control parameter. In general, the rank dependence of
$r_k^*$ is determined by the full distribution of reset distances and by
how the corresponding first-passage statistics are mixed within each
ordered arrival.

\section{Effect of the resetting protocol}
\label{sec:SI_protocol}

\begin{figure}[h]
	\centering
	\includegraphics[width=0.9\linewidth]{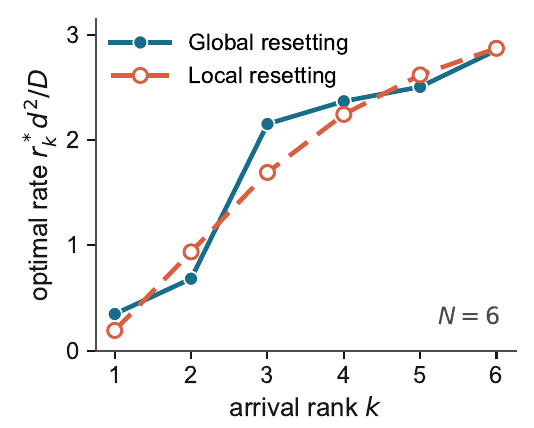}
	\caption{\textbf{Local and global resetting of non-interacting Brownian
			searchers.}
		Optimal resetting rate $r_k^*d^2/D$ versus arrival rank $k$ for $N=6$
		identical searchers under local and global resetting. Rates are expressed
		using the same per-particle resetting convention: under local resetting
		each particle has an independent clock of rate $r$, while under global
		resetting all particles are reset by a common clock of rate $r$.
		Both protocols retain the increase of the optimal rate with arrival rank,
		while the shared clock produces quantitative shifts through correlations
		between the particle trajectories.}
	\label{fig:protocol_SI}
\end{figure}
In the main text we focus primarily on local resetting, for which the first-passage times of non-interacting particles remain statistically independent. Here we compare this case with global resetting, in which a single Poisson clock resets all particles simultaneously. Although the particles do not interact physically in either case, the common reset times in the global protocol correlate their trajectories and therefore their ordered first-passage times.

To compare the two protocols, it is important to use a common rate convention. We denote by $r$ the resetting rate experienced by an individual particle. Thus, under local resetting each particle has an independent Poisson clock of rate~$r$, whereas under global resetting a single common Poisson clock of rate~$r$ resets all $N$ particles simultaneously. With this definition, the marginal resetting statistics of an individual particle are identical in both protocols; the difference lies entirely in the temporal correlations induced between searchers.

Figure~\ref{fig:protocol_SI} compares the optimal resetting rates for $N=6$ non-interacting Brownian searchers starting at the same distance $d$ from the target. The characteristic increase of $r_k^*$ with arrival rank~$k$ is preserved under both protocols. The shared reset clock nevertheless produces quantitative shifts in the optimal rates because the joint first-passage distribution no longer factorizes. For $k \geq 2$, these shifts are modest when the rates are expressed in the same per-particle convention. A much larger apparent separation between the two protocols results if the local rate is instead expressed as a total system-wide event rate, which differs by a factor of~$N$ from the corresponding per-particle rate.

Under local resetting, the ordered-passage statistics can be constructed directly from the independently restarted single-particle first-passage distributions, as described in the main text. Under global resetting this factorization is lost: all particles experience the same sequence of reset epochs, even though their Brownian motion between resets remains independent. The resulting correlations modify the quantitative location of the optimum but preserve the qualitative rank dependence observed for the homogeneous reference system. This is consistent with the distinction between local and global resetting established previously for the first arrival~\cite{BiroliMajumdarSchehr2023}.

These results motivate matching the resetting protocol when constructing the non-interacting baseline for each physical system considered in the main text.

\section{Matched Active Control for the Chemotactic System}
\label{sm:chemo-control}

The chemotactic system requires an active rather than Brownian non-interacting reference, as persistent self-propulsion and chemical coupling would otherwise enter the comparison simultaneously. We therefore construct a matched control consisting of $N=64$ non-chemotactic active particles incorporating the same self-propulsion speed $v_0$, rotational diffusion $D_R$, spatial confinement, target geometry, Poisson resetting rate, and boundary-return rules as the chemotactic simulations. The control retains active propulsion, rotational diffusion, confinement, target geometry, resetting, and boundary-return rules, while removing direct repulsion, chemotactic coupling, chemical deposition, and target-reinforcement dynamics.

The control simulations are conducted in a $32\times32\,\mu\mathrm{m}^2$ box. Both a stochastic Poisson reset and a collision with a non-target wall return a particle instantaneously to the origin with a uniformly randomized orientation. The confined single-particle process can thus be represented as a renewal sequence of independent excursions originating from $\mathbf{r}=\mathbf{0}$. We generate a pool of $4\times10^5$ excursions for each target distance by numerically integrating
\begin{equation}
	\dot{\mathbf r} = v_0\hat{\mathbf n}(\theta), \qquad \dot{\theta} = \sqrt{2D_R}\,\xi(t),
\end{equation}
with integration timestep $\Delta t = 0.05\,\mathrm{s}$, starting from $\mathbf r=\mathbf 0$ with isotropic initial orientations. For every excursion, we record its duration and whether it terminates at the target wall or at a non-target boundary.

For a given resetting rate $r$, each excursion is paired with an independent, exponentially distributed reset time. If the reset occurs before target arrival, a new excursion is initiated; if the target is reached first, the single-particle first-passage process terminates. Repeating this construction yields the single-particle first-passage distribution under resetting. The $N$-particle ordered-passage statistics then follow directly from binomial order statistics for identical independent searchers. Reusing the same excursion ensemble across all $r$ eliminates independent resampling noise between sampled resetting rates.

Including the finite box is essential for an absolute, calibration-free comparison. At $r=0$, $81.6\%$, $75.7\%$, and $65.0\%$ of individual excursions successfully reach the target for $L=2$, $3$, and $5\,\mu\mathrm{m}$, respectively, with median excursion durations of $105$, $193$, and $360\,\mathrm{s}$. Excursions encountering non-target walls are returned to the origin and renewed. Consequently, the confined control exhibits finite passage times even at $r=0$, predicting both absolute MKPTs and absolute optimal resetting rates without empirical scale factors. The chemotactic simulation output is reported in dimensionless form, with times in units of the rotational time $\tau_R = 1/D_R = 6.25\,\mathrm{s}$ and rates in units of $D_R = 0.16\,\mathrm{s}^{-1}$; all chemotactic values quoted here and in the main text have been converted to physical units using these factors.

Table~\ref{tab:chemo-control} summarizes the comparison between the matched control and the chemotactic simulations. At the last passage the chemotactic system is faster than the active control at all three target distances, by factors of $2.07$, $2.20$ and $2.32$. For example, at $L=3\,\mu\mathrm{m}$ the last-passage time falls from $2769.4\,\mathrm{s}$ in the control to $1257.5\,\mathrm{s}$ in the chemotactic system, while $r_{64}^*$ decreases from $0.016$ to $0.013\,\mathrm{s}^{-1}$.

\begin{table}[t]
	\centering
	\small
	\setlength{\tabcolsep}{3pt}
	\caption{\textbf{Matched confined active control for the chemotactic system.} Last-passage times are evaluated at $r=0$; optimal resetting rates are the discrete grid minima on the rate grid common to both datasets, accepted where they improve on the $r=0$ value by more than $1\%$ --- one rule applied identically to the control and to the chemotactic simulations, independent of the per-system extraction used for main-text Fig.~4f. All six entries are interior minima of the sampled grid, improving on $r=0$ by between $22\%$ and $67\%$; none is a bound. For $L=5\,\mu\mathrm{m}$, the control minimum is at approximately $0.00533\,\mathrm{s}^{-1}$, rounded to $0.005\,\mathrm{s}^{-1}$ in the table.}
	\label{tab:chemo-control}
	\vspace{6pt} % <-- Adds vertical space between caption and table content
	\begin{tabular}{cccccc}
		\hline\hline
		$L$ & $\langle T_{(64)}\rangle_{\mathrm{ctrl}}$ & $\langle T_{(64)}\rangle_{\mathrm{chem}}$ & $r^*_{64,\mathrm{ctrl}}$ & $r^*_{64,\mathrm{chem}}$ & \parbox[c]{1.1cm}{\centering $\frac{\langle T_{(64)}\rangle_{\mathrm{ctrl}}}{\langle T_{(64)}\rangle_{\mathrm{chem}}}$} \\
		($\mu\mathrm{m}$) & (s) & (s) & ($\mathrm{s}^{-1}$) & ($\mathrm{s}^{-1}$) & \\
		\hline
		2 & 2192.3 & 1060.6 & 0.023 & 0.023 & 2.07 \\
		3 & 2769.4 & 1257.5 & 0.016 & 0.013 & 2.20 \\
		5 & 4061.5 & 1748.2 & 0.005 & 0.006 & 2.32 \\
		\hline\hline
	\end{tabular}
\end{table}
\subsection{Rank-Resolved Effect of an Evolving Chemotactic Field}
\label{sm:chemo-speedup}

\begin{figure}[h!]
	\centering
	\includegraphics[width=0.8\linewidth]{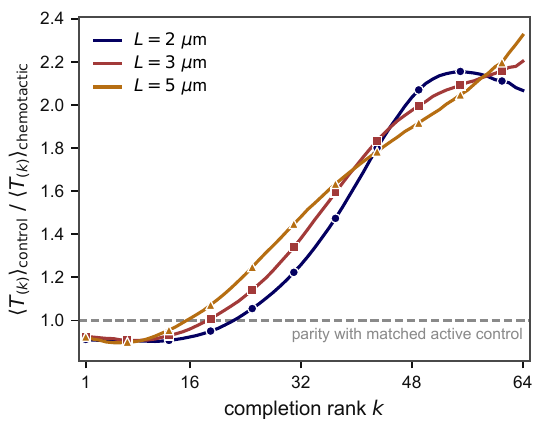}
	\caption{\textbf{Rank-resolved effect of chemotaxis relative to the matched confined active control.} Ratio $S_k = \langle T_{(k)}\rangle_{\mathrm{ctrl}} / \langle T_{(k)}\rangle_{\mathrm{chem}}$ at zero resetting ($r=0$) for target distances $L=2$, $3$, and $5\,\mu\mathrm{m}$; the dashed line marks parity with the control. Chemotaxis slightly delays the first arrival ($S_1 \approx 0.92$) and becomes beneficial only beyond $k=23$, $19$, and $16$ respectively, reflecting the progressive evolution and spatial organization of the collective field. The peak gains are $2.16$, $2.20$, and $2.32$. For $L=2\,\mu\mathrm{m}$ the ratio turns over at late ranks, whereas for $L=3$ and $5\,\mu\mathrm{m}$ it grows monotonically up to $k=N$. Markers are shown every six ranks for clarity.}
	\label{fig:S-speedup}
\end{figure}

The calibration-free control allows the effect of chemotaxis to be resolved as a function of arrival rank $k$. Figure~\ref{fig:S-speedup} plots the ratio evaluated at zero resetting ($r=0$),
\begin{equation}
	S_k = \frac{\langle T_{(k)}\rangle_{\mathrm{ctrl}}}{\langle T_{(k)}\rangle_{\mathrm{chem}}}.
\end{equation}
The effect changes sign along the arrival sequence. At the first arrival $S_1 \approx 0.91\text{--}0.92$, nearly independent of target distance $L$, so chemotaxis slightly \emph{delays} first discovery. Because all $64$ particles secrete chemoattractant from $t=0$, even the leading searcher moves through a collective field; that field channels the population onto shared routes at the expense of the broad spatial coverage that favours a first arrival. The sign of this early-rank effect agrees with the single-searcher result~\cite{Rudyak2025}, and the rank-resolved measurement adds that the penalty is confined to the earliest ranks.

$S_k$ then rises with rank and crosses unity at $k=23$, $19$ and $16$ for $L=2$, $3$ and $5\,\mu\mathrm{m}$, so the crossing moves to earlier ranks as the target recedes. We interpret this rank dependence as a direct consequence of the evolving chemoattractant landscape. As the search unfolds, particle trajectories progressively reinforce spatially preferred routes toward the target line. Consequently, later-arriving particles navigate a more strongly structured chemical field than the earliest arrivals. The gain reaches $S_k=2.16$ ($k=55$), $2.20$ ($k=64$) and $2.32$ ($k=64$), so the chemotactic system completes its last arrivals in roughly half the control time. For $L=2\,\mu\mathrm{m}$, $S_k$ turns over at the largest ranks, whereas for $L=3$ and $5\,\mu\mathrm{m}$ it increases monotonically up to $k=N$. The terminal ratio grows with target distance, $S_{64}=2.07$, $2.20$ and $2.32$.

The matched active control also settles whether chemotaxis advances the onset rank for beneficial resetting. This comparison is computed separately from the optima plotted in main-text Fig.~4f, under a single criterion applied identically to both datasets: the discrete minimum of $\langle T_{(k)}\rangle(r)$ on the rate grid common to the two systems, accepted only where it improves on the $r=0$ value by more than $1\%$, and recorded as $r_k^*=0$ otherwise. Neither the bootstrap nor the local quadratic fit of Sec.~\ref{sm:rstar_extraction} enters here. So defined, the control onset ranks are $k=27$, $27$ and $29$ for $L=2$, $3$ and $5\,\mu\mathrm{m}$, against $k=27$, $32$ and $35$ in the chemotactic system. Autochemotactic coupling therefore leaves the onset unchanged at the shortest target distance and \emph{delays} it by five and six ranks at the two larger ones: the chemical field sustains the unreset search for longer before restarting becomes worthwhile. The step-like staircase structure in $r_k^*$ reflects discrete rate grid resolution rather than intrinsic physical phase transitions.

\subsection{Geometric Assumptions}
\label{sm:chemo-assumptions}

The control relies on a square computational domain centered at the origin, $(x,y) \in [-16, 16]\,\mu\mathrm{m}$, a planar absorbing target line at $x=L$ spanning the box width, and instantaneous return to the origin upon contacting any non-target boundary. These boundary conditions uniquely determine the absolute passage times and speedup factors reported above; quantitative comparisons should be interpreted within the scope of this matched geometry.

\section{Extraction of optimal resetting rates}
\label{sm:rstar_extraction}

For each arrival rank $k$ the optimal resetting rate is
\begin{equation}
	r_k^* = \operatorname*{arg\,min}_{r \ge 0} \, \langle T_{(k)}(r) \rangle.
	\label{sm:eq:rstar}
\end{equation}
For the exact Brownian curves in the main text this is evaluated by
direct numerical minimization over $\log_{10} r$. For measured and simulated
data (main-text Fig.~4) a local quadratic $y = Ax^2+Bx+C$ is fitted in
$x = \log_{10} r$ around the discrete minimum, giving $r^* = 10^{-B/(2A)}$.

The fitting window and noise model differ by system. Colloids: seven points
centred on the minimum, weighted by $1/\sigma_i^2$ with
$\sigma_{\mathrm{eff}} = \sqrt{\mathrm{SEM}^2 + 1/12}$, the second term covering
integer-second rounding. ABPs: five points, with one noise scale from the
degree-of-freedom-corrected RMS residual of that fit. Chemotactic simulations,
which carry no reported uncertainties: one $\sigma$ per rank from pooled
residuals of sliding five-point quadratics along the whole curve. Where the
window would run off the sampled grid it is replaced rather than truncated --- by
the first seven points (colloids) or the last five (chemotactic), while for the
ABPs a window shorter than three points suppresses the fit and the discrete
minimum is reported.

Uncertainties come from a parametric bootstrap ($n_{\mathrm{boot}} = 4000$,
$y_i^{(b)} = y_i + \mathcal{N}(0,\sigma_i)$), refitting each draw and keeping the
vertex only if $A>0$ and it lies inside the window. The chemotactic fits add one
condition, that the value at the vertex fall below an independently perturbed
$r=0$ value, and re-locate the window on the minimum of each draw, which lets two
shallow competing basins exchange order; the colloid and ABP windows stay fixed
at the minimum of the unperturbed curve. Two acceptance criteria then apply in
sequence: the bootstrap distribution is used only if at least $20\%$ of draws
give interior minima, and the resulting median is reported as a finite $r_k^*$
(with the 16th and 84th percentiles as asymmetric bars) only if fewer than half
of all draws were rejected and the discrete minimum is not at either end of the
grid. Ranks failing either test are reported as bounds --- arrows with dashed
connecting segments for colloids and ABPs, and $r_k^*=0$ for the chemotactic
system, meaning no beneficial finite rate is resolved within the sampled range.

None of the three non-interacting controls is a direct minimization. The colloid
reference is a global-resetting Monte-Carlo calculation of the finite-$N$ order
statistics on the same rate grid, passed through the same seven-point weighted
bootstrap with $\sigma_i$ the Monte-Carlo standard error, and reported with the
symmetric half-width of the resulting interval. The ABP reference shares the
five-point window but is not bootstrapped; its uncertainty is the spread over
three simulation seeds. The confined non-chemotactic reference is not fitted at all: it is the discrete
grid minimum, kept only where it improves on $\langle T_{(k)}\rangle(0)$ by more
than $0.2\%$, which is why it is stepped. That rule produces the plotted profile;
the onset comparison quoted in the supplement is a separate calculation, in which
the same grid-minimum rule with a $1\%$ threshold is applied to the control and to
the chemotactic curves alike, so neither the bootstrap nor the quadratic fit enters
it.

Two conversions are applied after extraction, so plotted rates differ from those
in the source files: chemotactic rates are multiplied by
$\tau_R^{-1} = D_R = 0.16\,\mathrm{s}^{-1}$ (Eq.~\eqref{sm:eq:rstar} is invariant
under this rescaling), and ABP rates, stored as system-wide values, are divided
by $N=6$, so every panel shows a per-particle rate. Finally, for the
$L=4\,\mu$m colloidal series the four highest rates ($r\ge0.4\,\mathrm{s}^{-1}$)
are excluded because, at these highest resetting rates and this target distance,
the finite experimental duration yielded too few target arrivals to obtain
statistically reliable first-passage estimates.

\providecommand*{\mcitethebibliography}{\thebibliography}
\csname @ifundefined\endcsname{endmcitethebibliography}
{\let\endmcitethebibliography\endthebibliography}{}